\documentclass[conference]{IEEEtran}
\IEEEoverridecommandlockouts
\usepackage{cite}
\usepackage{amsmath,amssymb,amsfonts}
\usepackage{algorithmic}
\usepackage{graphicx}
\usepackage{breakcites}
\graphicspath{ {./images/} }
\usepackage{textcomp}
\usepackage{xcolor}
\usepackage{siunitx}
\usepackage{booktabs}
\usepackage{etoolbox}
\usepackage{multirow}
\usepackage{float}

  \renewrobustcmd{\bfseries}{\fontseries{b}\selectfont}
    \renewrobustcmd{\boldmath}{}
    \newrobustcmd{\B}{\bfseries}

\def\BibTeX{{\rm B\kern-.05em{\sc i\kern-.025em b}\kern-.08em
    T\kern-.1667em\lower.7ex\hbox{E}\kern-.125emX}}
\begin{document}

\title{Face Off: Polarized Public Opinions on Personal Face Mask Usage during the COVID-19 Pandemic\\

}

\author{\IEEEauthorblockN{\textsuperscript{} Neil Yeung, Jonathan Lai, Jiebo Luo}
\IEEEauthorblockA{
\textit{Department of Computer Science, University of Rochester}\\
\{nyeung, jlai11\}@u.rochester.edu, jluo@cs.rochester.edu}
}

% \IEEEoverridecommandlockouts
% \IEEEpubid{\makebox[\columnwidth]{978-1-7281-6251-5/20/\$31.00~\copyright2020 IEEE \hfill} \hspace{\columnsep}\makebox[\columnwidth]{ }}

\maketitle

% \IEEEpubidadjcol

\begin{abstract}
In spite of a growing body of scientific evidence on the effectiveness of individual face mask usage for reducing transmission rates \cite{mask_effective}, individual face mask usage has become a highly polarized topic within the United States. A series of policy shifts by various governmental bodies have been speculated to have contributed to the polarization of face masks. A typical method to investigate the effects of these policy shifts is to use surveys. However, survey-based approaches have multiple limitations: biased responses, limited sample size, badly crafted questions may skew responses and inhibit insight, and responses may prove quickly irrelevant as opinions change in response to a dynamic topic. We propose a novel approach to 1) accurately gauge public sentiment towards face masks in the United States during COVID-19 using a multi-modal demographic inference framework with topic modeling and 2) determine whether face mask policy shifts contributed to polarization towards face masks using offline change point analysis on Twitter data. First, we infer several key demographics of individual Twitter users such as their age, gender, and whether they are a college student using a multi-modal demographic prediction framework and analyze the average sentiment for each respective demographic. Next, we conduct topic analysis using latent Dirichlet allocation (LDA). Finally, we conduct offline change point discovery on our sentiment time series data using the Pruned Exact Linear Time (PELT) search algorithm. Experimental results on a large corpus of Twitter data reveal multiple insights regarding demographic sentiment towards face masks that agree with existing surveys. Furthermore, we find two key policy-shift events contributed to statistically significant changes in sentiment for both Republicans and Democrats.
\end{abstract}


\section{Introduction}

On March 11, 2020, the COVID-19 disease was officially declared a pandemic by the World Health Organization (WHO). As of August 19, the disease has infected over 22 million people worldwide with over 770,000 deaths. Individual face mask usage has been demonstrated to reduce transmission rates. However, within the United States, individual face mask usage has become a polarized topic.
Face masks can be differentiated into two categories. The first category are N-95 masks, which fit the National Institute for Occupational Safety and Health (NIOSH) certification for filtering at least 95 percent of airborne particles \cite{N95}. 
Only N-95 masks have been shown to effectively reduce both infection and transmission rates; thus, they are the only masks that healthcare workers can use to minimize the chance of contracting COVID-19 when interacting with patients within close quarters. Other types of masks include surgical grade masks, reusable masks, and homemade face covering; these masks have been shown to be effective at preventing transmission \cite{mask_effective}, but not at preventing a mask wearer from infection.
These masks are not NIOSH-Approved and are not sufficient for healthcare workers who are in close contact with patients on a daily basis.
During February, the Center for Disease Control and Prevention’s (CDC) official policy towards face masks was not to wear face masks for personal use. The CDC was concerned that recommending face mask usage as a policy would encourage hoarding behavior by non-healthcare workers and prevent healthcare workers from getting enough N-95 masks. These concerns were motivated by concerns of hoarding behavior\footnote{Hoarding behaviour was observed on other essential supplies, such as hand sanitizer}. 

On February 29, the Surgeon General tweeted:
\vspace{0.2in}
\begin{quote}
{\it 
    ``Seriously people- STOP BUYING MASKS! They are NOT effective in preventing general public from catching \#Coronavirus, but if healthcare providers can't get them to care for sick patients, it puts them and our communities at risk!"
    }
\end{quote}
\vspace{0.2in}

However, on April 03, the CDC reversed its earlier stance towards personal face mask usage, citing studies that showed the high amount of asymptomatic carriers of COVID-19 \cite{news2}. 
The rationale behind the policy change was that although there is no conclusive evidence that wearing a non-NIOSH-approved face mask prevents one from getting sick, asymptomatic carriers who wore non-N-95 face masks  prevent others from getting sick. Within the news media, shifts in governmental policy such as the one on April 3, along with inconsistent face mask policies by governmental officials, have been blamed for face masks becoming a political statement and an increasingly controversial topic \cite{news1}.
Given potential confusion about the effectiveness of masks as well as shifts in public policy towards face mask by governmental bodies such as the CDC, it becomes important to measure and understand both public perception of the public towards face masks and how public perception is affected as a result of policy shifts. To this end, our study utilizes demographic inference and sentiment analysis to track public perception of face masks. We also introduce change point analysis on the sentiment time series as an effective method of testing whether events cause statistically significant shifts in sentiment. 

In summary, our contributions in this study are as follows:

\begin{itemize}
    \item We employed state-of-the-art, multi-modal demographic inference techniques to characterize Twitter users over a wide amount of demographics of interest;
    \item We performed topic modeling and analysis on a novel and relevant subtopic of COVID-19---face mask usage; and 
    \item We applied iterative offline sentiment time series detection on Twitter data with demographic filters to confirm the findings from existing surveys and uncover new insights in public opinion on this polarizing national issue. 
\end{itemize}

\section{Related Work}
Twitter has been verified as a valuable source of data for analyzing and predicting various large-scale societal events such as elections \cite{1click}. During other pandemics such as the 2009 H1N1 pandemic \cite{H1N1} and the 2014 Ebola outbreak \cite{ebola}, Twitter has been used to monitor public sentiment towards the pandemics. Sentiment analysis and topic modeling research has been done on various aspects of COVID-19 such as investigating college students’ sentiment towards the pandemic \cite{Duong_college} and analyzing the use of controversial terms such as “Chinese Virus” \cite{chinese virus}. 

The majority of literature on face mask usage during COVID-19 has been on assessing face mask effectiveness and modeling effects of universal face mask usage \cite{mask_effective}, rather than measuring public perception towards face masks. Notably, there has not been any work specifically analyzing public perception towards face mask usage during COVID-19. Face mask usage is of particular interest considering recent news coverage surrounding politicization of mask usage and face mask mandates by corporations and state legislatures. Furthermore, demographic inference of users is important in light of sparse demographic reporting for those infected by COVID-19 and disparities found in infection and fatality rates along race, age, generational, racial, and economic lines \cite{covid_dem}.  We built upon previous research in social media demographic inference that analyzes users’ location, gender, race, and occupation using publicly available data \cite{M3, race}. Previous works on Twitter demographic inference differ from our approach in two ways: they either focus on a slim set of demographics (three or less) or only take a single aspect of a tweet object as input (only text, for example) \cite{twitter_demo}. 
In contrast, the recent advances in machine learning  makes it possible to utilize a \emph{multi-modal} approach to demographic inference along a \emph{wide} set of demographics.

Change point detection is an effective method to discover statistically significant shifts in time series data. Existing work on exploring sentiment time series graphs resulting from social media data has been sparse. Ref.  \cite{sent_change} uses online sentiment change detection to detect sentiment changes on Twitter streams of a specific hashtag and offline change detection is used to try to rediscover the shifts. Our approach differs in a few ways: we use an exact offline change point search algorithm rather than an approximate online change point search algorithm that detects changes in variance, an arguably more precise and efficient method \cite{PELT}; we iterate through multiple time series filtered through demographics (which is more specific than a hashtag constraint); and we use VADER \cite{VADER} for sentiment analysis which is valence-aware and specifically tuned for social media data, rather than a generic sentiment analyzer.

\begin{figure}[htpb]
\centerline{\includegraphics[width=.9\columnwidth]{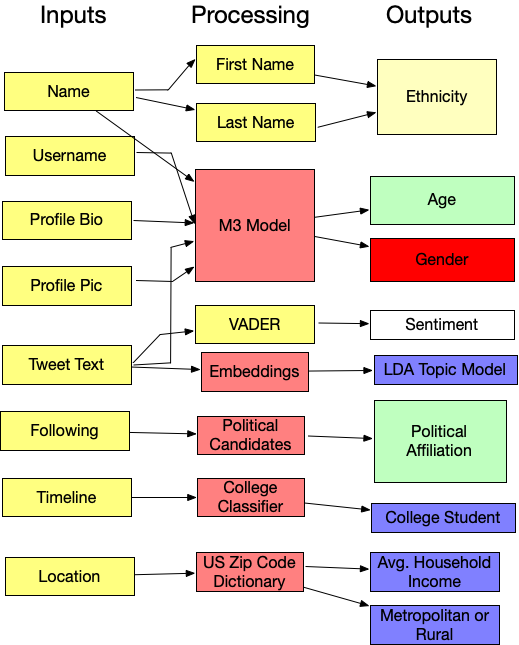}}
\caption{Overview of the demographic inference framework.}
\label{framework}
\end{figure}

\section{Demographic Analysis}
\subsection{Pre-processing}
We sampled tweets from a Covid-19 Twitter chatter dataset for scientific use \cite{dataset}. The dataset contains COVID-19 tweets that start from January 11, 2020 and, at the time of writing, continues to be gathered. We restricted our analysis of the data up to August 10, 2020. Each tweet is cleaned of URLs, hashtags, emojis, and  on Twitter. A version of the tweet containing emojis and hashtags is kept to feed into our sentiment analyzer. Every tweet is also tokenized, where each string is broken down individual words useful for sentiment analysis. A regex search for tweets that contain the terms in a dictionary of mask-related terms (e.g. ``face mask'' and ``mask'') along with a set of mask-related hashtags (e.g \#facemask) is applied to the data set and the tweet language is restricted to English. We recover the full JSON Twitter objects using hydration. Deleted tweets could not be recovered and are thus thrown out. Additionally, twitter accounts that were found to have a high chance of being an organizational entity rather than a person from the M3 model were also thrown out. After pre-processing, a total of 1.2 million tweets are gathered that mention mask-related terms spanning from January 11, 2020 to August 10, 2020.

\subsection{Valence-Aware Sentiment Analysis}
The Valence Aware Dictionary and Sentiment Reasoner (VADER) model is used for processing text and predicting valence scores and sentiment \cite{VADER}. The model utilizes a long short-term memory (LSTM) architecture and is trained on a gold standard of tweets created by 20 raters. The VADER model is chosen because it adjusts its rule-based compound score according to emojis and other common social media features that other sentiment models do not take account of. VADER outputs a number a normalized value from $[-1,1]$. Thus, we consider a positive VADER score of as having positive sentiment (such as $.04$), a negative VADER score as having negative sentiment (such as $-.04$), and a score of 0 as having neutral sentiment. When we refer to the VADER score of each demographic, we are referring to the average VADER score of that demographic. 

\subsection{Age, Gender, and College Student Classification}
Age and gender are two demographics of interest. In particular, Covid-19 has been shown to have a higher mortality rate amongst the older population and COVID-19 death data reveals that men suffer a higher fatality rate from COVID-19 \cite{covid_dem}. We extract age and gender using the M3 model, a multi-modal deep learning system for inferring the demographics of users from four sources of information from Twitter profiles: user’s name (first and last name in natural language), screen name (Twitter username), biography (short self-descriptive text), and profile image \cite{M3}. During the process of retrieving the profile images, we discard entries that contain a faulty profile image link, invalid picture format, deleted profiles, or users that do not output a prediction for any other reason. In the end, we are left with 1.09 million tweets which are able to have their age and gender predicted. 

\begin{figure}[htpb]
\centerline{\includegraphics[width=\columnwidth]{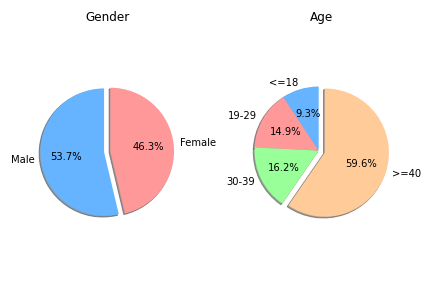}}
\caption{Age and Gender Distributions of All Users}
\label{ag_dist}
\end{figure}

College students are a demographic uniquely impacted by COVID-19. Previous work on analyzing college students has shown that college students react more negatively to COVID-19 than the general population \cite{Duong_college}. In the early days of the outbreak, it was reported that students continued to travel to popular spring break locations. We followed the methodology employed in \cite{Duong_college} to build a college student classifier. We gathered a gold standard balanced set of training data composed of 250 college students and 250 of non-college students. Two different raters hand labeled the entries. A score was assigned to each Twitter user if the raters agreed on the score. 

Ref. \cite{attributes} finds that users reveal their identifying attributes in the possessive form. We build upon a list of attributes that identifies a user as a college student from \cite{Duong_college} (e.g. ``my textbook'' and ``my professor'') and their point wise mutual information (PMI) to obtain a custom list of distinguishable attributes. We downloaded the full timelines of 100,000 users; due to Twitter rate limits, we are unable to download more than this number of timelines. We ran a 80/20 train-test split on our training data and converted the collection of timelines into a matrix of TF-ID features using a TF-ID vectorizer. We then extract relevant terms of the timelines for the training data and ran the relevant terms through a random forest classifier using our training labels to determine and predict class labels for whether the user is a college student or not using the list of identifying attributes. Users that displayed a high frequency of terms such as ``professor" and ``textbook" had their labels manually overridden to be college students. 

\begin{figure}[htpb]
\centerline{\includegraphics[width=\columnwidth]{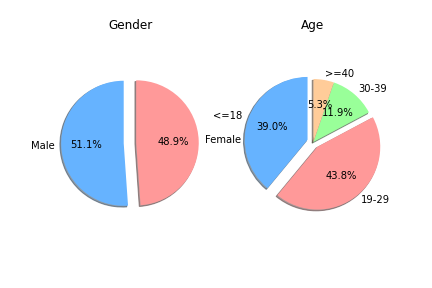}}
\caption{Age and Gender Distributions of College Students}
\label{col_ag_dist}
\end{figure}

\begin{figure}[htpb]
\centerline{\includegraphics[width=\columnwidth]{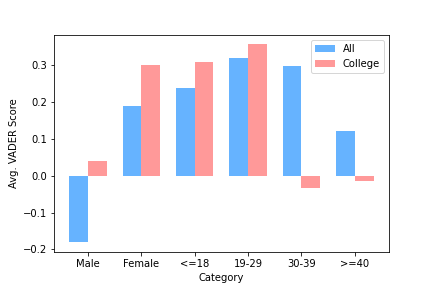}}
\caption{Average Sentiment Score across each Demographic}
\label{all_sent}
\end{figure}

A large majority of our data appears to be drawn from the Male, over 40 category when we look at all users. Previous Twitter demographic research \cite{attributes} have found that Twitter users skew young. However, a majority of our data finds Twitter users in the 40 and over age demographic (Fig. \ref{ag_dist}). One potential conclusion that can be drawn from this is that the 40 and over demographic disproportionately tweet about face masks as they are over represented. The gender distribution of the college students appears to coincide with the known fact that females attend college at a higher rate than males as the female college population is larger in the college student data set than the entire data set. The $\leq$18 and 19-29 categories make up $82.8\%$ of the college student data. This implies that the college student classifier is reasonably accurate, as most college students fall within this age range. 

Two key observations can be made on this data: females were found to have a higher average sentiment than males and college students have a different outlook on face masks than the general population. The fact that females have higher sentiment than males agrees with current surveys showing men are less likely to wear masks \cite{survey2}. College males had positive sentiment towards face mask in comparison to males in the overall data set. College females had higher sentiment than females in the overall data set. The college students in the $\leq$18 and 19-29 categories had higher sentiment than the counterpart in the complete data set. The college 30-39 and $\geq 40$ categories had opposite sentiments from their counterparts. We are not sure why this is. One factor that might have skewed the sentiment data for these two categories is the fact that the two categories should be underrepresented in the college population. Any inference that a user is both a college student and is over 30 is, according to probability, likely a result of an error from either classifier.

\subsection{Location and Average Household Income}
Approximately, 8.2\% (89,380) of the gathered twitter data which was able to have the user's age and gender inferred had a non-empty location. The location were checked against a list of cities within the United States. Only locations matching a city were kept. The median household income of the zip code associated with the city is retrieved and associated with each data point. We grouped household income based on the zip code's median household income's relationship to the real median household income within the entire United States (above, equal, below), which is 63,179 as of 2018 according to the U.S. Census Bureau. Each county is classified according to the 2013 Rural-Urban Continuum Codes (RUCC) classification scheme \cite{RUCC}. The RUCC classification scheme assigns each county in the county a number from 1-9. The schema differentiates metro areas based off population size and non-metro counties by level of urbanization. A county is assigned a number from 1 to 3 if it is considered a metro area and a number from 4 to 9 if it considered a non-metro area. We investigate the sentiment of each state through the four regions identified in the United States Census Bureau: Northeast, Midwest, South, and West. From this point forward, we refer to each region informally as East Coast, Midwest, South, and West Coast respectively.

Within our dataset, there are more metropolitan users than rural, more coastal users than the South and the Midwest, and more above average income users. This coincides with previous research on these specific demographic categories which implies that our methodology is reasonably accurate. 

\begin{figure}
    \centering
    \includegraphics[scale=1.5, width=\columnwidth]{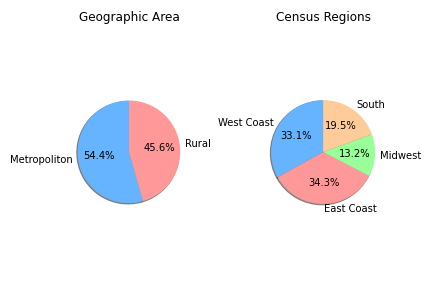}
    \caption{Geographic Area and Census Region Distribution.}
    \label{fig:cen_geo_dist}
\end{figure}

\begin{figure}
    \centering
    \includegraphics[width=\columnwidth]{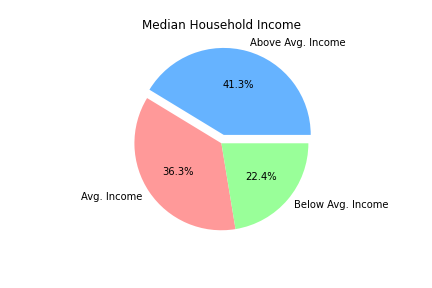}
    \caption{Median Household Income Breakdown}
    \label{fig:inc_dist}
\end{figure}

\begin{figure}
    \centering
    \includegraphics[width=\columnwidth]{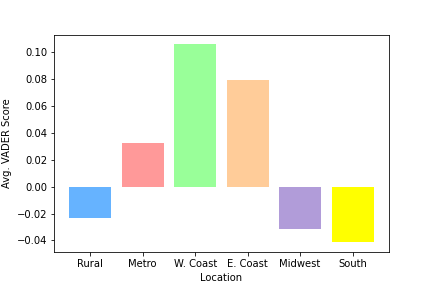}
    \caption{Average sentiment by geographic area and census region distribution.}
    \label{fig:geo_sent}
\end{figure}

\begin{figure}
    \centering
    \includegraphics[width=\columnwidth]{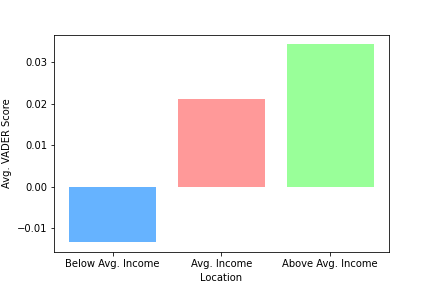}
    \caption{Average sentiment by Zip Code median household income.}
    \label{fig:inc_sent}
\end{figure}

Rural counties had more negative sentiment while urban counties had a higher positive sentiment. A possible explanation may be that face mask usage is more of a concern in metro environments due to higher population density in these areas. We also note that Twitter users are more represented in metropolitan and coastal regions in general. The Midwest and South also have negative sentiment in contrast to the coastal regions which have positive sentiment. This could potentially be due, again, to the population density difference, among other reasons (see Fig. \ref{fig:geo_sent} for comparisons). Above average income users and average users had positive sentiment, in contrast to below average income users. One explanation for this observed trend is that living in a zipcode with a higher median household income is correlated with living in a coastal, metropolitan region, as these regions tend to have a higher cost of living. Another explanation could be the fact the low-income households are more adversely affected by COVID-19. 

\subsection{Ethnicity}
We attempt to infer the ethnic profiles of each Twitter user in the 1.09 million users whose age and gender are previously inferred. First, we extract the first and last name utilizing a name parser; any middle names or titles are disregarded. We also craft a set of rules to remove any names that are not legitimate names (e.g. having a long sentence as a name). Then, the first and last names into an LSTM model trained on US census data from a Wikipedia database of names and their associated race \cite{race}. Ultimately, we are able to infer 900,000 legitimate names. For example, an entry with the first name John and the last name, “Smith,” would output a prediction for a Greater European, West European, British name. The ethnic prediction methodology yields 12 separate racial profiles corresponding with the Wikipedia training data. We group the 12 different racial profiles into 5 major groups: European (Germanic, Britain, East European), Hispanic (which includes Spanish origin names), African (as well as Muslim), East Asian and Indian. 

Notably, there appear to be less users with a name of European descent than expected based off existing census data. The most significant finding from the racial analysis is that the only ethnic profile to show negative sentiment is the European ethnic profile. This coincides with the ample media reports on the stronger resistance to face masks among this ethnic group \cite{survey1, survey2}.  Also, East Asian and Indian profiles have higher sentiment than African and Hispanic ethnic profiles. 

\begin{figure}
    \centering
    \includegraphics[width=\columnwidth]{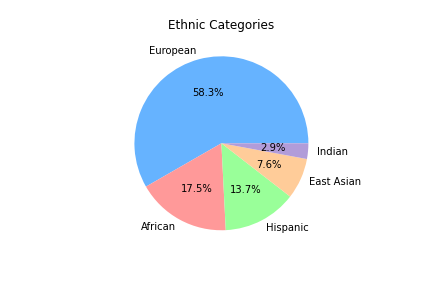}
    \caption{Ethnic profile distribution.}
    \label{ethc_dist}
\end{figure}

\begin{figure}
    \centering
    \includegraphics[width=\columnwidth]{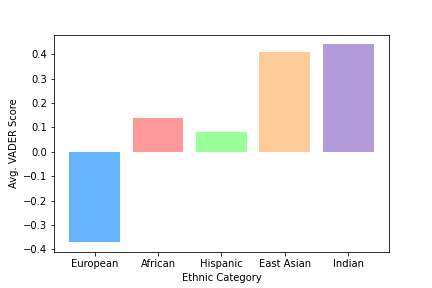}
    \caption{Sentiment by ethnic profile.}
    \label{ethc_sent}
\end{figure} 

\subsection{Political Affiliation}
Ref. \cite{1click} finds that 92\% of Twitter users who follow a political candidate, only follow candidates of a single party. The methodology is not able to be emulated exactly, due to the Twitter rate limits, so a modified methodology is utilized. First, the tweets are filtered for keywords drawn from a political keyword dictionary (which includes the names of presidential candidates, among others). This increases the likelihood that the user is politically active (e.g. follows a Democrat politician or Republican politician). Next, the ``following'' list of each user that passes the initial filter is downloaded. The timeline is checked against a list of the verified senators and presidential candidate Twitter account ids. Per the 92\% one party heuristic from \cite{1click}, it is assumed that if a user follows a candidate of one party, they only follow candidates of that party, and the respective party affiliation is attached to the user. This allows for retrieval of a higher volume of users with assigned political affiliation given Twitter rate limits. Ultimately, 241,199 users are able to be assigned a political affiliation.

\begin{figure}
\includegraphics[width=\columnwidth]{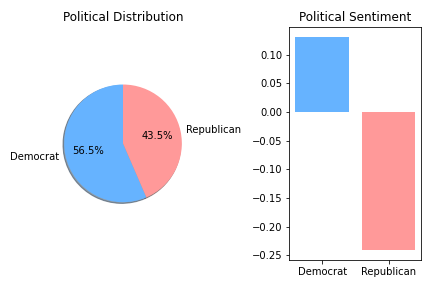}
\caption{Sentiment and distribution across political affiliation. }
\end{figure}

The fact that there are more Democrats within the collected data is unsurprising, given that existing Twitter demographic research shows that most Twitter users are identify as liberal \cite{twitter_demo}. It is notable that Republicans have sentiment that is more negative than the Democrats' sentiment is positive. This finding also agrees with current surveys \cite{survey2}.

\begin{figure*}
\includegraphics[width=\textwidth]{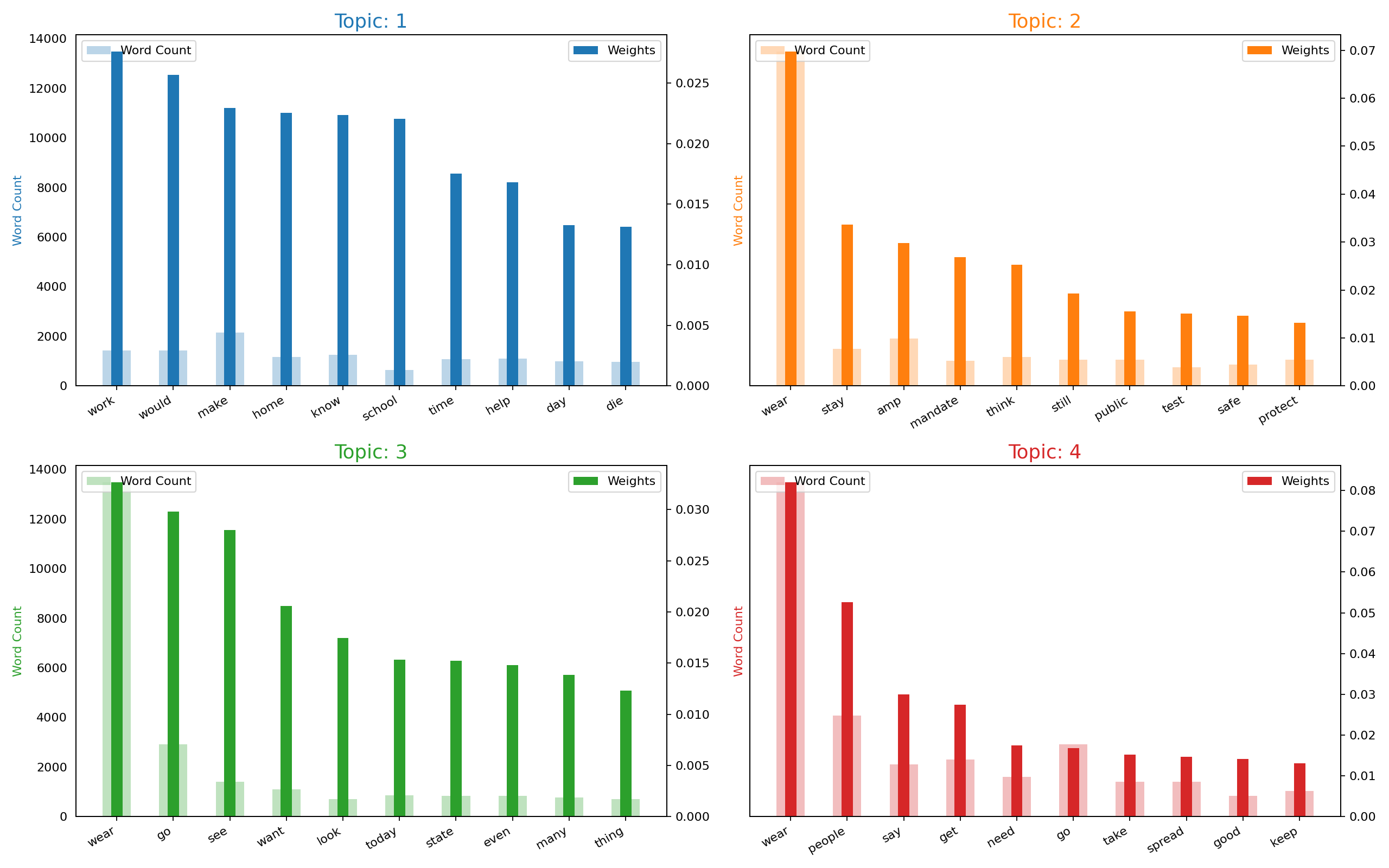}
\caption{Word count and importance of topic keywords.}
\end{figure*}

\section{Topic Analysis}
Latent Dirichlet Allocation (LDA) is utilized to discover universal topics within the chatter dataset. Only noun, verb, adjective, and adverb lemmas are kept in each token to extract the most relevant information. The bigrams and trigrams are computed. The \emph{NLTK} stop words package is supplemented with a custom dictionary of virus-related words (i.e. infection, covid19) to ensure only novel topics. The LDA model with the highest combination of coherence score and subjective interpretability is chosen. Four topics are chosen to analyze and table \ref{tab:table1} shows the rough topic groupings with an example representative tweet.
\begin{table}[ht!]
  \begin{center}
    \caption{Latent Dirichlet Analysis Topics Description and Example Tweets}
    \label{tab:table1}
    \begin{tabular}{l|l}
    \hline
      \textbf{Topic Description} & \textbf{Example Tweet} \\
      \hline
      \multirow{5}{.3\columnwidth}{Topic 1: Mask at School and Work} & No requirements appear to be included\\ & for masks or any other additional\\ & safety and prevention. When the majority \\ & of teachers and support staff are \\ &  home sick thatll be a short school year. \\
      \hline
      \multirow{6}{.3\columnwidth}{Topic 2: Governmental Mask Policy} & About time Governor DeSantis mandates  \\ & the wearing of masks in public. \\ & Lives before politics Governor,\\ & and for people who refuse to wear masks, \\  & idiosyncrasies have no place in a pandemic. \\& I applaud the Florida lawyer for his efforts. \\
      \hline
      \multirow{2}{.3\columnwidth}{Topic 3: Post-Pandemic Events} & Wear a dang mask   if you want to \\ & see Big Tex again. \\
      \hline
      \multirow{6}{.3\columnwidth}{Topic 4: Public Health Concerns} & ``From a medical standpoint it would be \\ & extremely rare to have a condition where \\ & you truly could not wear a mask,'' \\ & said Dr. Clay Callison, a pulmonary \\ & disease doctor and and UT Medical \\ & Center's Chief Medical Information Officer. \\
      \hline
    \end{tabular}
  \end{center}
\end{table}

% or any other additional safety and prevention \\ When the majority of teachers &amp; support staff are home sick \\ thatll be a short school year} \\

\subsection{Analysis}

Each topic is interpreted to have an overall theme. Subtopics are identified by reading through representative example tweets and then listing subtopics that reoccur often. Four topics are identified: Masks at School and Work, Governmental Mask Policy, Post-Pandemic Events, and Public Health Concerns.

The first topic talk about masks at school and work. Two subtopics are highlighted: concern for a lack of safety precautions in school and places of work work and observations about mask usage at school and work. One interesting reoccurring concern within the first subtopic is that places of work and school are reopening prematurely, which is expressed within the example tweet in Table \ref{tab:table1}. Many tweets express doubt towards proposed efforts to reopen safely. Tweets within this subtopic may claim that ``teachers and students will die" when schools reopen. The second subtopic of tweets contains observations about the rates of mask usage at school and work. The observations almost always center on a lack of face mask usage at school and work. For example, a few tweets talk about seeing crowded schools in a state where no students are wearing masks.

The second topic talks about governmental mask mandates and policies. Many of the tweets talk about state officials such as the state governor or other elected public officials. Notably, a large portion of the tweets talk about Trump's usage of masks. There are many tweets that comment on the tweet Trump created on July 20, 2020 about wearing a mask. The tweets either claim that Trump is being hypocritical or thanking Trump for being patriotic (per the language of Trump's tweet). Some tweets express a conspiracy that Trump is wearing a mask because he is being ``forced to'' by some entity. When discussing political officials other than Trump, tweets range from congratulating state officials for their policies towards face masks or complaining about the state officials' policy. The topic also includes discussion about mask mandates at stores such as Walmart, Walgreens, and Kroger. There is also many mentions of the CDC and WHO. For example, one tweet claims that the CDC is guilty of ``criminal negligence'' regarding ``lies about mask''. This verifies our earlier hypothesis that Twitter users may have been confused about policy changes. 

The third topic talks about post-pandemic events. The tweets talk about what people plan to do when the pandemic is over or what events that they missed due to the pandemic. Some of the tweets are formulated as a plea to wear masks so that people can go to certain events that they are not able to during quarantine. For example, a tweet may ask people to wear masks ``if they ever want to see a concert again''. When examining tweets early in the pandemic, from around late March to early May, there are tweets talking about how the pandemic would be over soon. For example, a tweet may talk about how a year from now, everyone would look back and say, ``remember when we had to wear masks''. 

The fourth topic talks about general public health concerns. Under the umbrella of general public health concerns, there is a recurrent line of discussion on the effectiveness of masks. Tweets may cite research on masks or claim that masks are not effective at all. Tweets that claim that masks are not effective cite multiple reasons. A few cite religious protection; for example, one tweet says that there is ``no need to wear a mask'' and that one can ``JUST PRAY''. A few tweets mention N-95 masks. The tweets that mention N-95 masks appear to display understanding of the difference between N-95 masks and other masks. For example, a tweet that talks about N-95 masks may mention an N-95 mask's ability to protect one ``to the microns''. A non-trivial amount of tweets are about self-made masks. These tweets detail masks being made out of miscellaneous materials such as cloth.

\section{Experiment}
We are interested in studying the shifts in public policy by governmental entities as these shifts have been speculated to have contributed to the polarization surrounding individual mask usage. We assume that a shift in policy enacts a shift in the sentiments of tweets about masks. Specifically, we are interested to discover if any public policy shift has a statistically significant effect on the sentiment time series. Since we have access to the full sentiment time series, our experiment can be formulated as an \emph{offline} change point detection problem. Broadly, our methodology is the following: we iteratively run an offline change point detection algorithm to detect any statistically significant shifts in average sentiment for each of the sentiment time series filtered by each demographic. Then, we look within the news for a governmental shift event $\pm 3$ days around the detected change point to account for reactions to news streaming in for a few days and associate the news event with the change point. 
\subsection{Problem Formulation}
Modifying notation from \cite{PELT}, we formulate our problem. Let $y=\{y_t\}_{t=1}^{N}$ be a sentiment time series and let any sentiment time series be denoted as $\mathcal{T} \subset N$. We create a sentiment time series for each demographic $D$ as follows: set each $y_t$ as the mean of all sentiments $s$ of the users $u \in D$ in that demographic are included in the calculation.

A change point is a time $\tau$ in a finite time series $\mathcal{T}$ such that there is a statistically significant change in variance $\sigma$ between $y_{1:t}$ and $y_{t+1:\mathcal{T}}$. There are $K$ such points where $K \geq 0$. We utilize the Pruned Exact Linear Time (PELT) algorithm which assumes we have unknown $K$, have access to the full time series $\mathcal{T}$, runs in $\mathcal{O}(N)$, and looks for the best time series segmentation $\mathcal{T}$ that solves exactly for

\begin{IEEEeqnarray}{c}
\min _{\mathcal{T}} V(\mathcal{T})+\beta(\mathcal{T})
\end{IEEEeqnarray}

where $\beta>0$ is a smoothing parameter and $V(\mathcal{T})$ is a criterion which we seek to minimize\footnote{For more detail on PELT, please read \cite{PELT}}.

\begin{figure}[htpb]
\centerline{\includegraphics[width=\columnwidth]{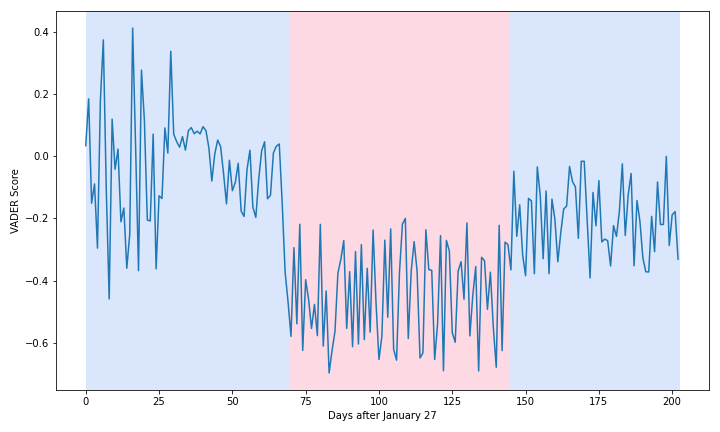}}
\caption{Republican sentiment time series change point discovery with Pelt search.}
\label{rep_time}
\end{figure}

\begin{figure}[htpb]
\centerline{\includegraphics[width=\columnwidth]{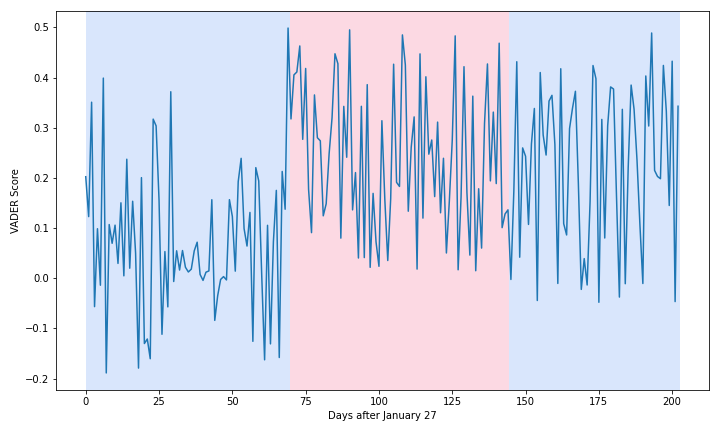}}
\caption{Democrat sentiment time series change point discovery with Pelt search.}
\label{dem_time}
\end{figure}

We run PELT on every demographic sentiment time series as well as the sentiment time series of the complete data. We only find one proper segmentation: within time series is filtered by political affiliation; the results are displayed in Fig. \ref{rep_time} and \ref{dem_time}). PELT finds two change points: at $66$ and $145$ days after January 27, 2020. This roughly this corresponds to April 04, 2020 and July 20, 2020. A change in the color of the background of the graph (alternating blue and red) denotes a detected change point. 
\subsection{Discussion}
When looking at the news events relevant to masks that might have caused the sentiment change, two policy shift events could potentially be correlated with the change: the shift in policy by the CDC on April 03 (1 day before April 04) and Trump's tweet on July 20, 2020 about wearing a mask. The specific tweet that President Trump sent from his account said:
\vspace{0.2in}
\begin{quote}
{\it 
    ``We are United in our effort to defeat the Invisible China Virus, and many people say that it is Patriotic to wear a face mask when you can’t socially distance. There is nobody more Patriotic than me, your favorite President!"
    }
\end{quote}
\vspace{0.2in}

During the previous topic analysis, these two events were cited by many tweets. Only the time series filtered by political affiliation yielded results with PELT search. From January 27 to April 03, both Republicans and Democrats had positive average sentiment. However, after April 03, Republican average sentiment dropped to negative sentiment while Democrat average sentiment became positive. After Trump tweets about wearing a mask on July 20, 2020, there was a shift up in Republican average sentiment. However, Republican average sentiment stayed negative. Democrats, in contrast, had a downshift of average sentiment, but stayed positive in sentiment score. This finding is interesting, as it may imply that political affiliation is the only consistent demographic in predicting how a person's perception will change towards a mask policy change, even though political affiliation is correlated with some of the other demographics inferred in this study (such as a rural location). This finding builds upon existing survey data that show Democrats are more likely to wear masks than Republicans \cite{survey2}.

\subsection{Limitations}
The limitation of this approach is an inability to treat for confounders. Any detected change point may have resulted from another news event that is not the CDC shift or Trump's tweet about wearing a mask. However, given extensive news media coverage on these two events, there is a high likelihood that any statistically significant sentiment shift detected in those date ranges is a result of the major events. 

\section{Conclusion and Future Work}
Consistent public policies are essential for managing public health. To this end, we introduced multi-modal demographic inference combined with valence-aware sentiment analysis topic analysis. We also conduct an experiment utilizing offline change point detection for sentiment analysis.

Our demographic inference finds some illuminating insights: 1) males had negative average sentiment while females had positive sentiment towards face masks, 2) the same age and gender of college students, when compared to the broader population, had different perception towards face masks, 3)
people with European ethnic profiles had negative average sentiment while every other ethnic profile had positive sentiment.

Finally, our experiment using PELT search for Twitter sentiment analysis shows that the April 03, 2020 CDC policy shift event, and Trump tweeting about wearing a mask on July 20, 2020, when considering the time series filtered by political affiliation demographic, were statistically significant sentiment change points. We also find that Republicans appeared to decrease their average VADER score in reaction to the CDC policy shift, while increasing their average VADER score in reaction to Trump's tweet. Democrats, on the other hand, had their average VADER score increase in reaction to the CDC policy shift and decrease in reaction to Trump's tweet.

Potential future work could include the use of a filtering algorithm to reliably catch fake twitter accounts. This would decrease the amount of bots and fake twitter accounts within the data set and increase the number of authentic users in our data set. We acknowledge that some demographics may prove difficult to predict. As one example, many Asian Americans have European names that they prefer to use when online. Future modifications to address this deficiency may involve utilizing a multi-modal approach to predict ethnicity, similar to how age and gender is predicted.

Social media data mining and sentiment analysis have the potential to be important tools in measuring public response to abrupt policy changes during pandemics. More governments and health organizations can leverage social media data mining technology to improve policy outcomes.

\section{Acknowledgment}
We thank Viet Dong for providing the student classification algorithm.


\begin{thebibliography}{99}

\bibitem{mask_effective} D.K. Chu et al., ``Physical distancing, face masks, and eye protection to prevent person-to-person transmission of SARS-CoV-2 and COVID-19: a systematic review and meta-analysis,'' \emph{The Lancet}, vol. 395, issue 10242, pp.1973-198, Jun. 2020.

\bibitem{N95} Centers for Disease Control and Prevention. ``NIOSH-Approved N95 Particulate Filtering Facepiece Respirators'', \emph{CDC}. [Online]. Available: https://www.cdc.gov/niosh/npptl/topics/respirators/disp\_part/n95list1.html. [Accessed: May 5, 2020]. 

\bibitem{n95eff} J.D. Smith et al., ``Effectiveness of N95 respirators versus surgical masks in protecting health care workers from acute respiratory infection: a systematic review and meta-analysis.” vol. 188, no. 8, pp. 567-574, May 2020. [Online]. Available: NBCI, https://www.ncbi.nlm.nih.gov/pmc/articles/PMC4868605/. [Accessed: Aug 10, 2020].

\bibitem{dataset} J. M Banda et al., \emph{A large-scale COVID-19 Twitter chatter dataset for open scientific research - an international collaboration}. [Dataset], Zenodo. Available: http://doi.org/10.5281/zenodo.3831406. [Accessed: August 2, 2020].

\bibitem{sent_change} S. K. Tasoulis, A. G. Vrahatis, S. V. Georgakopoulos and V. P. Plagianakos, ``Real Time Sentiment Change Detection of Twitter Data Streams,'' in \emph{2018 Innovations in Intelligent Systems and Applications (INISTA)}. Available: IEEE Xplore, http://www.ieee.org. [Accessed: Aug. 10. 2020].

\bibitem{VADER} C.J. Hutto and E. Gilbert, ``VADER: A Parsimonious Rule-based Model for Sentiment Analysis of Social Media Text,'' in \emph{Eighth International Conference on Weblogs and Social Media}, Ann Arbor, MI, June 2014.

\bibitem{M3} Z. Wang et al., ``Demographic Inference and Representative Population Estimates from Multilingual Social Media Data,'' in \emph{Proceedings of the 2019 World Wide Web Conference}, May 2019.

\bibitem{attributes} 
B. Shane, and B. Van Durme, ``Using conceptual class attributes to characterize social media users,'' in \emph{Proceedings of the 51st Annual Meeting of the Association for Computational Linguistics}, Aug. 2013.

\bibitem{H1N1} C. Chew and G. Eysenbach, ``Pandemics in the age of Twitter: content analysis of Tweets during the 2009 H1N1 outbreak,'' \emph{PloS one}, Nov. 2010. [Online]. Available: doi:10.1371/journal.pone.0014118.

\bibitem{ebola} EHJ Kim et al., ``Topic-based content and sentiment analysis of Ebola virus on Twitter and in the news,'' in \emph{Journal of Information Science}, Vol 42, Issue 6, 2016.

\bibitem{PELT} C. Truong, L. Oudre, and N. Vayatis, ``Selective Review of Offline Change Point Detection Methods,'' in \emph{Signal Processing}, 167:107299, 2020.

\bibitem{RUCC} J. Cromartie, ``Rural-Urban Continuum Codes”. United States Department of Agriculture Economic Research Service, 2013. [Online]. Available: https://www.ers.usda.gov/data-products/rural-urban-continuum-codes/documentation [Accessed May 1, 2020]. 

\bibitem{race} G. Sood and S. Laohaprapanon, ``Predicting Race and Ethnicity From the Sequence of Characters in a Name'', May 2019, arXiv:1805.02109.

\bibitem{chinese virus} L. Chen, H. Lyu, T. Yang, Y. Wang and J Luo, ``In the eyes of the beholder: Sentiment and topic analyses on social media use of neutral and controversial terms for covid-19,''April 2020, arXiv preprint arXiv:2004.10225.

\bibitem{Duong_college} V. Duong, P. Pham, T. Yang, Y. Wang, and J. Luo,``The Ivory Tower Lost: How College Students Respond Differently than the General Public to the COVID-19 Pandemic,'' April 2020, arXiv preprint arXiv:2004.09968.

\bibitem{twitter_demo} A. Mislove et al, ``Understanding the Demographics of Twitter Users'', in \emph{Proceedings of the Fifth international AAAI Conference on Weblogs and Social Media}, 2011.

\bibitem{covid_dem} J.B. Dowd et al., ``Demographic science aids in understanding the spread and fatality rates of COVID-19'', in \emph{Proceedings of the National Academy of Sciences}, vol. 117, no. 18, pp. 9696-9698, 2020. Available: 10.1073/pnas.2004911117. [Accessed 16 August 2020].

\bibitem{flu} H. Achrekar, A. Gandhe, R. Lazarus, S. Yu and B. Liu, ``Predicting Flu Trends using Twitter data,'' in \emph {2011 IEEE Conference on Computer Communications Workshops (INFOCOM WKSHPS)}, Shanghai, 2011, pp. 702-707, doi: 10.1109/INFCOMW.2011.5928903.

\bibitem{1click} Y. Wang, X. Zhang, and J. Luo, ``When Follow is Just One Click Away: Understanding Twitter Follow Behavior in the 2016 Presidential Election'', arXiv, abs/1702.00048, 2017. 

\bibitem{survey1} Katz, J., Sanger-Katz, M. and Quealy, K., ``A Detailed Map Of Who Is Wearing Masks In The U.S.,'' \emph{The New York Times} [Online] Available: https://www.nytimes.com/interactive/2020/07/17/upshot/coronavirus-face-mask-map.html [Accessed 15 September 2020].            
\bibitem{survey2} M. Brenan, ``Americans' Face Mask Usage Varies Greatly by Demographics,'' Jul. 13, 2020, \emph{Gallup}. [Online]. Available: https://www.nytimes.com/interactive/2020/07/17/upshot/coronavirus-face-mask-map.html [Accessed 15 September 2020].            

\bibitem{news1} R. Rojas, ``Masks Become a Flash Point in the Virus Culture Wars,” \emph{The New York Times}, 3 May 2020. [Online]. Available: www.nytimes.com/2020/05/03/us/coronavirus-masks-protests.html.

\bibitem{news2} A. Karni and M. Astor, ``As Leaders Urge Face Masks, Their Behavior Muffles the Message", \emph{The New York Times}, 2020. [Online]. Available: https://www.nytimes.com/2020/04/22/us/politics/coronavirus-masks.html. [Accessed: 15- Aug- 2020].

\end{thebibliography}
\end{document}